\documentclass[a4paper,12pt]{article}            
\usepackage{jheppub}
\pdfoutput=1

\usepackage[utf8]{inputenc}
\usepackage{amsmath}
\usepackage{mathtools}
\usepackage{mathrsfs}  
\usepackage{tensor}
\usepackage{bbm}
\usepackage{multirow}
\usepackage{enumitem}
\usepackage[normalem]{ulem}
\allowdisplaybreaks
\usepackage{comment}
\usepackage{doi}

\title{Dissipative electrically driven fluids}
\author[1,2]{Andrea Amoretti,}
\author[1,2]{Daniel K. Brattan,} 
\author[1,2]{Luca Martinoia,}
\author[1,2]{Jonas Rongen.}

\emailAdd{andrea.amoretti@ge.infn.it}
\emailAdd{danny.brattan@gmail.com}
\emailAdd{luca.martinoia@ge.infn.it}
\emailAdd{jonas.ludovico.rongen@edu.unige.it}

\affiliation[1]{Dipartimento di Fisica, Universit\`a di Genova,
via Dodecaneso 33, I-16146, Genova, Italy}
\affiliation[2]{I.N.F.N. - Sezione di Genova, via Dodecaneso 33, I-16146, Genova, Italy}

\begin{abstract}
{\ We consider entropy generating flows for fluids that achieve a steady state in the presence of a driving electric field. Having chosen one among the space of stationarity constraints that define such flows we show how energy and momentum relaxation are related in the presence of dissipation. Furthermore, we find that if such a fluid obeys Onsager reciprocity then the incoherent conductivity must be identically zero and consequently makes no contribution to the observable AC or DC charge conductivities.}
\end{abstract}
\makeatletter
\gdef\@fpheader{}
\makeatother
\begin{document}

\maketitle

\section{Introduction}\label{sec:Intro}

{\noindent Drude's model of electron transport provides the pedagogical framework for describing charge transport in a conductor under the influence of an externally applied electric field \cite{AshcroftMermin}. According to the model, to prevent the indefinite acceleration of charge carriers driven by the electric field - which continuously supplies energy and momentum - a mechanism that dissipates energy and momentum is required. Consequently the energy and momentum sinks balance the electric field and the system achieves a steady state.}

{\ However, in the standard hydrodynamic description of a charged fluid in an external electric field, the hydrostatic conditions indicate that a stationary state is achieved when the external electric field $\mathbb{E}$ is balanced by the gradient of the chemical potential $\mu$. Hence, unlike in Drude's model, the fluid's velocity and the external electric field are treated as independent degrees of freedom. To address this issue, we previously investigated how to modify the hydrostatic constraints in a boost-agnostic, electrically driven fluid through the incorporation of relaxation terms \cite{Amoretti:2022ovc}. This allows the fluid to exhibit Drude-like behaviour, where the electric field in these stationary states is not entirely counteracted by the gradient of the chemical potential. This approach necessitates that the fluid velocity becomes itself a thermodynamic variable, introduced as a chemical potential conjugate to momentum. As a result, different inertial frames represent distinct hydrodynamic states and can no longer be related to each other through boost transformations. This contrasts with boost-invariant hydrodynamics, where one can always move to a frame with zero spatial velocity, effectively removing it as a variable. The framework for describing these types of fluids is known as ``boost-agnostic'' hydrodynamics  \cite{deBoer:2017ing,deBoer:2017abi,deBoer:2020xlc,Armas:2020mpr,Novak:2019wqg}. We will review the fundamentals of this approach shortly.}

{\ As in Lagrangian mechanics, we introduced the energy and momentum relaxation terms — analogous to non-conservative forces — by hand to the equations of motion after having derived the conservative components using a variational principle. To maintain consistency, we assume that we can separate the relaxation contributions into two types: those that can be expressed in terms of stationary tensor structures and those that vanish at stationarity. In reference \citep{Amoretti:2022ovc}, we focused on the first type, while this work aims to extend the analysis by examining the relation of the energy and momentum relaxation terms of the latter type. This means we will consider flows that achieve a steady state under the influence of an external driving electric field and that include tensor structures in the constitutive relations that are first-order in derivatives and vanish at stationarity. Consequently, we will also encompass entropy-generating flows.}

{\ Let us briefly comment on the magnitude of the relaxation terms. If the momentum relaxation rate $\Gamma\sim\tau^{-1}$ is very small, it can be treated as a minor correction to the hydrodynamic conservation laws. In such cases, either the relaxation term restricts the space of stationary solutions to the hydrodynamic equations of motion or it changes them. The former case is followed in \cite{Amoretti:2023vhe}. In \cite{Amoretti:2022ovc} and in this work, we consider the latter regime where $\Gamma$ is small enough (compared to microscopic decay modes) for momentum to remain a relevant (quasi-)conserved charge, however it enters in a way that stationary conditions are modified. In this case the relaxation terms are comparable to other thermodynamic quantities, and thus cannot be accounted for by simply adding small corrections to the conservation equations. Finally, in the case where $\Gamma$ is large and momentum is strongly decaying, momentum ceases to be a useful hydrodynamic variable and becomes irrelevant for dynamics at large scales.}

{\ This can be compared to the standard treatment of an external magnetic field \cite{Amoretti:2020mkp,Amoretti:2021lll,Amoretti:2022acb,Amoretti:2022vxq,Amoretti:2014kba}. When the magnetic field is small enough, $B\sim\mathcal{O}(\partial)$, it simply modifies the momentum conservation equation. In the limit of a very strong magnetic field, $B\gg T^2$, the system transitions out of the hydrodynamic regime. For intermediate values, $B\sim\mathcal{O}(1)$, the magnetic field becomes a crucial part of the thermodynamics and explicitly alters the constitutive relations, giving rise to non-dissipative Hall terms. Similarly, we can consider $\Gamma\sim\mathcal{O}(\partial^2)$ to follow the approach taken in \cite{Hartnoll:2007ih,Amoretti:2023vhe}, while this work focuses on the case where $\Gamma\sim\mathcal{O}(\partial)$.
}

{\ We will now proceed to the main part of this paper as follows: In section \ref{sec:boostagnosticreview} we provide a brief review of the formulation for a boost agnostic, charged, non-dissipative electrically driven fluid in the absence of relaxation terms. For this we geometrise the fluid's thermodynamics by coupling it to an Aristotelian spacetime \cite{deBoer:2020xlc,Armas:2020mpr,Amoretti:2022ovc} and provide the hydrostaticity conditions that determine stationary fluid flows. We also sketch the computation of the constitutive relations associated with these stationary configurations, with further details available in \cite{Amoretti:2022ovc}. In sections \ref{sec:boostzero} and \ref{sec:boostone} we move away from absolute conservation of energy and momentum by introducing relaxation terms into the hydrodynamic equations of motion at order zero and one, respectively. More specifically in section \ref{sec:boostone} we identify the first-order derivative tensor structures in the constitutive relations that vanish at stationarity. By enforcing positivity of entropy production, we constrain the relevant transport coefficients and characterise energy relaxation in terms of momentum relaxation. We then apply our theory in section \ref{sec:conductivity}, where we compute the thermo-electric conductivities of our system in the absence of microscopic time-reversal invariance. Additionally, we identify the part of the charge current that is neither carried by heat nor spatial momentum, defining a generalisation of the incoherent current.  In section \ref{sec:onsager}, we then impose Onsager reciprocity and show that the AC thermo-electric conductivities assume their Drude form.  Moreover, we find that the contribution from the incoherent conductivity to the electric conductivity vanishes. Finally, we conclude with a discussion of our results in section \ref{sec:discussion}.}

\section{Brief review of non-dissipative boost agnostic hydrodynamics}
\label{sec:boostagnosticreview}

{\noindent In the following section we will briefly review the formalism for describing a $(d+1)$-dimensional, charged, boost agnostic fluid in the presence of an external electric field $\vec{\mathbb{E}}$ \cite{Amoretti:2022ovc,deBoer:2017ing,Armas:2020mpr}. We will not introduce any relaxation terms in this section.}

{\ In global thermodynamic equilibrium, we take the fluid to be described by a pressure $P$ which is a function of temperature $T$, chemical potential $\mu$, spatial velocity $\vec{v}$ and the external electric field $\vec{\mathbb{E}}$. Deriving the pressure with respect to these source terms yields various thermodynamic densities:
\begin{subequations}
\begin{align}
\label{Eq:Densities}
n =&\left( \frac{\partial P}{\partial \mu} \right) \; ,~~\rho_{\mathrm{m}} = 2 \left( \frac{\partial P}{\partial \vec{v}^2} \right) \; ,~~  s = \left( \frac{\partial P}{\partial T} \right) \; , ~~ \\
& \beta_{\mathbbm{P}} = \left( \frac{\partial P}{\partial (\vec{\mathbbm{E}} \cdot \vec{v})} \right) \; , ~~
\kappa_{\mathbbm{E}} =  2 \left( \frac{\partial P}{\partial \vec{\mathbbm{E}}^2} \right) \; ,
\end{align}
\end{subequations}
where $s$ is the entropy density of the fluid, $\rho_{\mathrm{m}}$ is the kinetic mass density and $n$ the particle number density. The parameters $\kappa_\mathbb{E}$ and $\beta_\mathbb{P}$ have been introduced in \citep{Amoretti:2022ovc} and are related to the momentum density $\vec{P}$ and polarisation density $\vec{\mathbbm{P}}$ by
\begin{subequations}
\begin{align}
		 \label{Eq:momentumdef}
		\vec{P} =& \left( \frac{\partial P}{\partial \vec{v}} \right)  = \rho_{\mathrm{m}} \vec{v} + \beta_{\mathbbm{P}} \vec{\mathbbm{E}}   \; , \\
		\label{Eq:polarisationdef}
		 \vec{\mathbbm{P}} =& \left( \frac{\partial P}{\partial \vec{\mathbbm{E}}} \right) = \kappa_{\mathbbm{E}} \vec{\mathbbm{E}} + \beta_{\mathbbm{P}} \vec{v} \; .
 \end{align}
\end{subequations}
We will also assume that our fluid satisfies an Euler relation of the form~\cite{Amoretti:2022ovc}
\begin{align} \label{Euler relation}
	sT &=  \epsilon + P - \mu n - \rho_m \vec{v}^2 - \kappa_{\mathbb{E}} \vec{\mathbb{E}}^2 - 2 \beta_{\mathbb{P}} \vec{\mathbb{E}} \cdot \vec{v}\nonumber\\
	&=\epsilon+P-\mu n-\vec{\mathbb{P}}\cdot\vec{\mathbb{E}}-\vec{P}\cdot\vec{v}
\end{align}
with $\epsilon$ the energy density.}

{\ Given the above global thermodynamic equilibria, we seek to describe a situation where our system of interest can be broken up into patches of local thermodynamic equilibrium and the flows between them. A precise way to achieve this is to geometrise the thermodynamics of the fluid. For fluids that break boost invariance, this geometrisation naturally couples the fluid to an Aristotelian spacetime which consists of a manifold equipped with a clock one-form $\tau_\mu$ and a spatial metric $h_{\mu \nu}$ with signature $(0,1,...,1)$. This should be compared with a Lorentzian geometry where the clock-form and spatial metric combine to form a locally Lorentz boost invariant metric $g_{\mu \nu} = - \tau_\mu \tau_\nu + h_{\mu \nu}$.}

{\ We further assume that the spatial metric can be decomposed into vielbeins\footnote{Since our spacetime is Aristotelian the vielbeins $e^a_\mu$ transform under $SO(d)$.}
\begin{equation}
	h_{\mu \nu} = \delta_{a b} e^a_{\mu} e^b_{\nu}, 
\end{equation}
with $a,b = 1,2,..., d,$ and $\mu = 0,...,d$ where $d$ denotes the number of spatial dimensions. The square matrix $(\tau_\mu, e^a_{\mu})$ then has a non-vanishing determinant denoted by
\begin{equation}
	\label{Eq:Dete}
	e = \text{det}(\tau_\mu, e^a_\mu).
\end{equation}
Consequently, as the determinant is non-vanishing, the matrix is invertible with its inverse decomposing into $(-\nu^\mu, e^\mu_a)$. The components of this inverse satisfy various useful relations:
\begin{equation}
	\nu^\mu \tau_\mu=-1, \quad \nu^\mu e_\mu^a=0, \quad e_a^\mu \tau_\mu=0,
	\quad e_a^\mu e_\mu^b=\delta_a^b \; ,
\end{equation}
and allow us to define a preferred symmetric two-tensor $h^{\mu \nu}$ by
\begin{equation}
	h^{\mu \rho} h_{\rho \nu} = \delta^\mu_\nu + \nu^\mu \tau_\nu. 
\end{equation}
Although this tensor is not the inverse of the spatial metric $h_{\mu \nu}$, it can be interpreted as such on the spatial hypersurfaces defined by the clock-form $\tau_\mu$.}

{\ Let us remark at this point for the sake of completeness that connections that are compatible with the independent one-form and spatial two-tensor of Aristotelian spacetime, i.e.
\begin{equation}
	\nabla_\mu \tau_\nu = 0,\quad
	\nabla_\mu h_{\nu \rho} = 0
\end{equation}
are not unique, as opposed to the pseudo-Riemannian spacetime, where the unique metric compatible and torsion free connection is the Levi-Civita connection.  This however will not be important for us as, in the end, we will be interested in flat Aristotelian spacetime in Cartesian coordinates (henceforth denoted FSCC). In this limit the geometry-characterising tensors reduce to
\begin{equation} \label{FSCC}
	\begin{split}
		& \tau_\mu = \delta^0_\mu, \quad h_{\mu \nu} = \delta^i_\mu \delta^j_\nu \delta_{ij}, \quad \nu^\mu = - \delta^\mu_0, \quad h^{\mu \nu} = \delta^{\mu}_i \delta^{\nu}_j \delta^{ij}, \\
		& \partial_\mu \tau_\nu = 0, \quad \partial_\mu h_{\nu \rho} = 0, 
	\end{split}
\end{equation}
where Latin indices only run over spatial directions.}

{\ Now that we have discussed the geometry of Aristotelian spacetimes sufficiently, we can proceed with describing fluids moving on them. We begin by determining the fluid's thermodynamic quantities in terms of the geometric quantities $\{ \tau_\mu, h_{\mu \nu}, A_{\mu} \}$, where the gauge field accounts for the presence of a chemical potential and external electric field. The flow of a stationary fluid will be described by a Killing vector field $\beta^{\mu}$ which satisfies:
\begin{equation} \label{hydrostatic constraints}
	\begin{aligned}
		\mathcal{L}_\beta \tau_\mu & =0, \\
		\mathcal{L}_\beta A_\mu & =0, \\
		\mathcal{L}_\beta h_{\mu \nu} & =0 .
	\end{aligned}
\end{equation}
In this way $\beta^\mu$ - the thermal vector - can be thought of as the time-direction of the fluid's dynamical evolution; that is to say it generates time translations.}

{\ Given any such time-like Killing vector field we can define a notion of temperature and chemical potential using loops around the thermal circle \cite{Jensen:2012jh}. In terms of our Aristotelian tensors we find
\begin{subequations}
\label{thermodynamic frame}
\begin{equation}
		T =  \frac{1}{\tau_\mu \beta^\mu}, \qquad
		\mu =  T \left( A_\mu \beta^\mu + \Lambda \right),
\end{equation}
where $\Lambda$ is a gauge parameter present to ensure that $\mu$ remains gauge invariant. With these definitions to hand we further define the fluid velocity in terms of the thermal vector to be
\begin{equation}
	\beta^\mu = \frac{u^\mu}{T} \; .
\end{equation}
Finally, the electric field is given by
\begin{equation}
	F_{\mu \nu} = 2 \partial_{[\mu} A_{\nu]} = \mathbb{E}_\mu \tau_\nu - \mathbb{E}_\nu \tau_\mu
\end{equation}
\end{subequations}
which, given it is expressed in terms of the gauge invariant quantity $F_{\mu \nu}$, is naturally gauge invariant.}

{\ As is discussed in \cite{Jensen:2012jh} we can employ the definitions (\ref{thermodynamic frame}) to re-express the Killing constraints (\ref{hydrostatic constraints}) in terms of the hydrodynamic quantities. However, in what follows, we will only need the hydrostaticity conditions for flat Aristotelian spacetime in Cartesian coordinates:
\begin{subequations}
\label{conditions hydrostatic equilibrium}
\begin{equation}
	\label{conditions hydrostatic equilibrium 1}
	\begin{array}{lll}
		\partial_\mu T=0, & \quad \partial_t v^i=0, & \quad \partial_i v_j+\partial_j v_i=0\\
		\partial_i \mathbb{E}_j-\partial_j \mathbb{E}_i=0, & \quad \partial_t \mathbb{E}_i+v^j \partial_j \mathbb{E}_i+\mathbb{E}_j \partial_i v^j=0, &
	\end{array}
\end{equation}
and
\begin{equation} \label{Eq:ModifiedHydrostaticity}
	 \mathbb{E}_i - \partial_i \mu = 0 \; .
\end{equation}
\end{subequations}
The general curved spacetime reformulations can be found in  \cite{Amoretti:2022ovc}. Importantly, these relations \eqref{conditions hydrostatic equilibrium} tell us how a fluid can flow while still being stationary - one solution of which on flat space is global thermodynamic equilibrium.}

{\ Thus far we have geometrised our thermodynamics and discussed conditions for the existence of stationary fluid flows \eqref{hydrostatic constraints} even when the spacetime is curved. Now let $W[\tau,h,A]$ be the generating functional for the correlation functions of such flows. This generating functional can be expanded order by order in derivatives. For example, the leading term is
\begin{align}
    \label{Eq:ConstitutiveGenerator}
    W_{(0)}[\tau,h,A]&=\int d^{d+1}x\; e \; P(T,\mu, h^{\mu \nu} \mathbbm{E}_{\mu} \mathbbm{E}_{\nu}, h_{\mu \nu} u^{\mu} u^{\nu} , \mathbbm{E}_{\mu} u^{\mu})
\\
&\equiv \int d^{d+1}x\; e\; P\left(T,\mu,\vec{\mathbbm{E}}^2,\vec{v}^2,\vec{v} \cdot \vec{\mathbbm{E}} \right)\nonumber
\end{align}
where $P$ is the thermodynamic pressure. From the generating functional $W$ we can define the following important one-point functions:
	\begin{eqnarray}
		\label{Eq:ChargeDefs}
		\frac{\delta W}{\delta h_{\mu \nu}} = \frac{e}{2} T^{\mu \nu} \; , \qquad
		\frac{\delta W}{\delta \tau_{\mu}} = - e T^{\mu} \; , \qquad
		\frac{\delta W}{\delta A_{\mu}} = e J^{\mu} \; .
	\end{eqnarray}
Diffeomorphism and gauge invariance of $W$ then ensure that these one-point functions satisfy the following \cite{Amoretti:2022ovc} constraints
\begin{subequations}
\label{Eq:HydroEqns}
\begin{align}
	& e^{-1} \partial_{\mu} \left( e T\indices{^\mu_\rho} \right) + T^{\mu} \partial_{\rho} \tau_{\mu} - \frac{1}{2} T^{\mu \nu} \partial_{\rho} h_{\mu \nu} - F_{\rho \mu} J^{\mu} = 0 , \label{stress energy tensor equation} \\
	& e^{-1} \partial_{\mu} \left( e J^{\mu} \right) = 0. \label{eom current}
\end{align}
where $J^{\mu}$ is the $U(1)$ charge current and $T\indices{^\mu_\nu}$ is a stress tensor defined by
	\begin{eqnarray} \label{definition energy-momentum tensor}
		T\indices{^\mu_{\nu}} = - T^{\mu} \tau_{\nu} + T^{\mu \rho} h_{\rho \nu} \; .
	\end{eqnarray}
\end{subequations}
The equations \eqref{Eq:HydroEqns} are no more than conservation of energy/momentum and charge in an Aristotelian spacetime and they form the equations of motion of hydrodynamics. To complete the hydrodynamic description we must also supply constitutive relations expressing these currents in terms of the thermodynamic charges. It is an important consequence of the compatibility between the Killing conditions \eqref{hydrostatic constraints} for stationary flows of the fluid and the above discussed symmetries that the hydrodynamic equations \eqref{Eq:HydroEqns} are automatically satisfied for stationary flows.}

\section{Relaxation at order zero} \label{sec:boostzero}

{\noindent Having geometrised the fluid in the absence of relaxation, we now move away from absolute conservation of the stress-energy-momentum tensor complex to allow for relaxation. We add by hand non-conservative forces in the natural and normative way one would in mechanical systems. In particular, these forces may be viewed as constraints on the stationary flows of the system. In the FSCC limit, while maintaining $U(1)$ charge conservation, the relaxed equations of motion then take the form
	\begin{subequations}
	 \label{relaxed EOM FSCC}
	\begin{eqnarray}
		\partial_t \varepsilon+\partial_i J_{\varepsilon}^i-\mathbb{E}_i J^i &=& - \hat{\Gamma}_{\varepsilon}, \\
		\partial_t P_i+\partial_j T \indices{^j_i} - n \mathbb{E}_i &=&  - \hat{\Gamma}^{i}_{\vec{P}},\\
		\partial_t n+\partial_i J^i &=& 0 \; ,
	\end{eqnarray}
\end{subequations}
where $\epsilon = - T \indices{^0_0}$ denotes the energy density, $J^i_\epsilon = - T \indices{^i_0}$ the energy current,  $P_i = T \indices{^0_i}$ the spatial momentum density, $n$ the charge/number density and $\hat{\Gamma}_{\varepsilon}$ and $\hat{\Gamma}^{i}_{\vec{P}}$ the energy and momentum relaxation terms in flat spacetime. The LHS of \eqref{relaxed EOM FSCC} is nothing more than \eqref{Eq:HydroEqns} expressed in FSCC. In the following we parametrize the momentum relaxation in terms of order-zero vectors as $\hat\Gamma^i_{\vec{P}}=n\left(\Gamma_{\vec{P}}P^i+\Gamma_{\mathbb{\vec{P}}}\mathbb{P}^i\right)$.}

{\ We follow the typical procedure in hydrodynamics and solve the (non-)conservation equations \eqref{relaxed EOM FSCC} order by order in a derivative expansion. This requires us to be precise about the derivative order counting for the electric field. Implicitly, as it entered the global thermodynamics, we took $\vec{\mathbbm{E}}$ to be order zero above. Consequently, the polarization $\vec{\mathbb{P}}$ is of order zero \cite{Amoretti:2022ovc,Kovtun:2016lfw} and the (non-)conservation equations of motion at lowest order in derivatives, after imposing all stationarity conditions \eqref{conditions hydrostatic equilibrium} other than \eqref{Eq:ModifiedHydrostaticity}, become\footnote{Notice we introduced an extra factor of $n$ on the RHS compared to \cite{Amoretti:2022ovc} for presentation purposes.}
\begin{equation}
	\label{Eq:OrderZeroEqns}
	\begin{split}
		& n v^i \left( \mathbb{E}_i - \partial_i \mu \right) = \hat\Gamma_\varepsilon+ \mathcal{O}(\partial), \\
		& n \left( \mathbb{E}_i - \partial_i \mu \right) = n\left(\Gamma_{\vec{P}} P_i + \Gamma_{\vec{\mathbb{P}}} \mathbb{P}_i\right) + \mathcal{O}(\partial). 
	\end{split}
\end{equation}
In the absence of relaxation terms, one must take the gradient of the chemical potential to be order zero in derivatives \cite{Amoretti:2022ovc,Kovtun:2016lfw} if the electric field is strong (order zero in derivatives). On the other hand, in the presence of relaxation terms, there exists at least two regimes where we can satisfy \eqref{Eq:OrderZeroEqns}. The first of these is that \eqref{Eq:ModifiedHydrostaticity} is maintained and consequently both the momentum and energy relaxations must vanish at stationarity. This subsequently imposes a constraint on the thermodynamics of local equilibrium through the zeroes of the relaxation terms \cite{Toner:1998,Amoretti:2024obt}.}

{\ Alternatively, we can assert that neither the LHS nor the RHS of \eqref{Eq:OrderZeroEqns} are zero on their own, but instead these expressions must be treated as conditions for hydrostaticity as a whole. The hydrostaticity condition \eqref{Eq:ModifiedHydrostaticity} is modified to
	\begin{eqnarray}
		\label{Eq:ModifiedHydrostaticity2}
		\mathbb{E}_i - \partial_i \mu =  \Gamma_{\vec{\mathbb{P}}} \mathbbm{P}_i + \Gamma_{\vec{P}} P_i \; ,
	\end{eqnarray}
and consequently, one finds that the energy and momentum relaxations are related
\begin{equation} \label{relaxation relation}
	\hat\Gamma_\varepsilon = \left( \Gamma_{\vec{\mathbb{P}}} \mathbbm{P}^i + \Gamma_{\vec{P}} P^i \right)n v_{i} \; .
\end{equation}
This is what one expects from the Drude model. One can subsequently show that there is no entropy production in the relaxed case at the ideal order even in the presence of the relaxation terms if \eqref{relaxation relation} holds.}

{\ We now make two important simplifications for the rest of the paper; firstly, we shall assume that we are supplied an exact hydrostaticity condition. In particular, we will not allow it to be modified order by order in derivatives and the resultant equations of motion must respect this constraint at all orders in derivatives. Adding such additional terms does not in principle pose any difficulties. Secondly, because of the rapid growth in the number of transport terms, it will generally be convenient to work with an electric field $\vec{\mathbb{E}}$ at order one in derivatives rather than order zero. Similarly, we take the derivative of the chemical potential and the relaxation terms to be order one. Given these constraints we take our hydrostaticity condition to be:
\begin{equation}
	\begin{split}
		\label{Eq:FinalHydrostaticityConstraint}
		\mathbb{E}_i - \partial_i \mu - \Gamma P_i = 0 \; ,
	\end{split}
\end{equation}
at all orders in derivatives.}

{\ While our new hydrostaticity constraint \eqref{Eq:FinalHydrostaticityConstraint} holds at all orders in derivatives, it does not follow that \eqref{relaxation relation} is without derivative correction as it was a derived consequence of the equations of motion on hydrostatic solution. In the following we would like to understand how this constraint between the relaxation terms changes once we begin to consider entropy producing flows.}

\section{Relaxation at order one}
\label{sec:boostone}

{\noindent We will require our fluid to locally obey the second law of thermodynamics such that the divergence of the entropy current $S^{\mu}$ is non-negative
\begin{equation} \label{divergence entropy current}
	e^{-1} \partial_{\mu} \left( e S^{\mu} \right) \geq 0,
\end{equation}
where $e = \text{det}(\tau_\mu, e^a_\mu)$ and where the entropy current is the most general expression built from the fluid variables that reduces to $s u^\mu$ for vanishing derivative terms. Rewriting this divergence will allow us to isolate the non-hydrostatic contributions to the constitutive relations in what follows. In particular, we can separate the non-entropy producing non-hydrostatic corrections, from those that produce entropy.}

{\ We define the canonical entropy current by covariantising the thermodynamic Euler relation (\ref{Euler relation}). This gives the following expression\footnote{At the end of section~\ref{sec:boostagnosticreview} we have decided to take $\vec{\mathbb{E}}$ to be order one in derivatives in writing the constitutive relations, however here we are still considering it to be order zero since it does not cause too many complications.}
\begin{equation} \label{covariantised entropy current}
	S^{\mu}_{\text{can}} = - T\indices{^\mu_\nu} \beta^{\nu} + P \beta^{\mu} - \frac{\mu}{T} J^{\mu} - \kappa_{\mathbb{E}} \mathbb{E}^{\nu} \mathbb{E}_{\nu} \beta^{\mu} - 2 \beta_{\mathbb{P}} \mathbb{E}^{\nu} u_{\nu} \beta^{\mu} \; ,
\end{equation}
whose divergence is 
\begin{equation} \label{entropy divergence}
	\begin{split}
		e^{-1} \partial_{\mu} \left( e S^{\mu}_{\text{can}} \right) = & - e^{-1} \partial_{\mu} \left( e T\indices{^\mu_\nu} \beta^{\nu} \right) - e^{-1} \partial_{\mu} \left( e J^{\mu} \frac{\mu}{T}  \right) + e^{-1} \partial_{\mu} \left(e P \beta^{\mu} \right) \\
		& - e^{-1} \partial_{\mu} \left( e \kappa_{\mathbb{E}} \mathbb{E}^{\nu} \mathbb{E}_{\nu} \beta^{\mu} \right) - e^{-1} \partial_{\mu} \left( e 2 \beta_{\mathbb{P}} \mathbb{E}^{\nu} u_{\nu} \beta^{\mu} \right) \; .
	\end{split}
\end{equation}
This expression, as written, contains all orders in the derivative expansion of the hydrodynamic variables.}

{\ The first two terms of the RHS of (\ref{entropy divergence}) can be rewritten using the (non-)conservation equation of the energy-momentum tensor (\ref{stress energy tensor equation}) and the conservation equation of the charge current (\ref{eom current}). In particular, we can covariantise the energy and momentum relaxation terms and modify \eqref{stress energy tensor equation} to
\begin{equation} \label{eom SEM tensor}
	\begin{split}
		e^{-1} \partial_{\mu} \left( e T\indices{^\mu_\rho} \right) + T^{\mu} \partial_{\rho} \tau_{\mu} - \frac{1}{2} T^{\mu \nu} \partial_{\rho} h_{\mu \nu} = F_{\rho \mu} J^{\mu} + \Gamma_\rho \; .
	\end{split}
\end{equation}
Contracting this equation with the thermal vector $\beta^{\mu}$ leads to
\begin{equation} \label{contraction}
	\begin{split}
		e^{-1} \partial_{\mu} \left(e \beta^{\rho} T\indices{^\mu_\rho} \right) = & \, T\indices{^{\mu}_\rho} \partial_{\mu} \beta^{\rho} - T^{\mu} \beta^{\rho} \partial_{\rho} \tau_{\mu} + \frac{1}{2} T^{\mu \nu} \beta^{\rho} \partial_{\rho} h_{\mu \nu} + F_{\rho \mu} \beta^{\rho} J^{\mu} + \beta^{\rho} \Gamma_\rho \; ,
	\end{split}
\end{equation}
which can be further manipulated so that it is expressed in terms of the hydrostatic constraints \eqref{hydrostatic constraints}. One subsequently obtains
\begin{equation} \label{rewritten eom SEM}
	e^{-1} \partial_{\mu} \left(e \beta^{\rho} T\indices{^\mu_\rho} \right) = - T^{\mu} \mathcal{L}_{\beta} \tau_{\mu} + \frac{1}{2} T^{\mu \nu} \mathcal{L}_\beta h_{\mu \nu} + F_{\rho \mu} \beta^{\rho} J^{\mu} + \beta^{\rho} \Gamma_\rho
\end{equation}
such that the divergence of the canonical entropy current takes the form
\begin{equation}\label{entropy divergence 2} 
	\begin{split}
		e^{-1} \partial_\mu\left(e S_{\mathrm{can}}^\mu\right) = & T^\mu \mathcal{L}_\beta \tau_\mu-\frac{1}{2} T^{\mu \nu} \mathcal{L}_\beta h_{\mu \nu} - J^\mu \left( \mathcal{L}_\beta A_\mu - \partial_\mu \left( \frac{u^\nu A_\nu - \mu}{T} \right) \right) \\
		& +e^{-1} \partial_\mu\left(e P \beta^\mu\right) -e^{-1} \partial_\mu\left(e \kappa_{\mathbb{E}} \mathbb{E}^\nu \mathbb{E}_\nu \beta^\mu\right) -e^{-1} \partial_\mu\left(e 2 \beta_{\mathbb{P}} \mathbb{E}^\nu u_\nu \beta^\mu\right) \\ 
		& - \beta^{\rho} \Gamma_\rho \; .
	\end{split} \;
\end{equation}
}

{\ Let us introduce the subscript $_{(0)}$ to indicate ideal fluid terms. The ideal fluid part of the entropy current $S^\mu_{(0)}$ has zero divergence, $e^{-1} \partial_\mu \left( e S^{\mu}_{(0)} \right) = 0$, due to thermodynamic identities and hence does not contribute to entropy production. Using this observation we obtain the following identity
\begin{equation} \label{ideal entropy divergence2}
	\begin{split}
		e^{-1} \partial_{\mu} & \left( e  \left\{ P \beta^{\mu} - 2 \beta_{\mathbb{P}} \mathbb{E}^{\nu} u_{\nu} \beta^{\mu} - \kappa_{\mathbb{E}} \mathbb{E}^{\nu} \mathbb{E}_{\nu} \beta^{\mu} \right\} \right) \\
		= & - T^{\mu}_{(0)} \mathcal{L}_{\beta} \tau_{\mu} + \frac{1}{2} T^{\mu \nu}_{(0)} \mathcal{L}_{\beta} h_{\mu \nu} + J^{\mu}_{(0)} \mathcal{L}_{\beta} A_{\mu} - \partial_{\mu} \left( \beta^{\rho} A_{\rho} \right) J^{\mu}_{(0)} + \partial_{\mu} \left(\frac{\mu}{T} \right) J^{\mu}_{(0)} \; .
	\end{split}
\end{equation}
which allows us to eliminate the pressure from \eqref{entropy divergence 2}. We arrive to
\begin{eqnarray}
		\label{Eq:FirstOrderPartEntropy0}
		e^{-1} \partial_{\mu} \left( e S_{\mathrm{can}}^{\mu} \right) &=  \, \left(T^{\mu} - T^{\mu}_{(0)} \right) \mathcal{L}_\beta \tau_\mu - \frac{1}{2} \left(T^{\mu \nu} - T^{\mu \nu}_{(0)} \right) \mathcal{L}_\beta h_{\mu \nu} - \left( J^{\mu} - J^{\mu}_{(0)} \right) \delta_\mathcal{B} A_\mu \nonumber \\
		& - \beta^{\rho} \Gamma_\rho,
\end{eqnarray}
where we have introduced the notation
\begin{equation}
	\begin{split}
		\delta_\mathcal{B} A_\mu & \coloneqq \mathcal{L}_\beta A_\mu - \partial_\mu \Lambda = \mathcal{L}_\beta A_\mu - \partial_\mu \left( \frac{u^\nu A_\nu - \mu}{T} \right)
	\end{split}
\end{equation}
In the absence of relaxation $\Gamma_{\rho}=0$, upon imposition of the hydrostaticity constraints \eqref{hydrostatic constraints}, we see that \eqref{Eq:FirstOrderPartEntropy0} vanishes identically, independently of the derivative order at which we are working. This is to be expected, if entropy was produced on hydrostatic flows then the system would not be stationary.}

{\ If we include relaxation terms in \eqref{ideal entropy divergence2}, the Killing constraint $\delta_{\mathcal{B}} A_{k} = 0$ needs to be adjusted so that for hydrostatic flows we find no entropy production. We saw that in the FSCC limit, denoted by a tilde, the condition for hydrostaticity is modified to
\begin{equation}
	\tilde{\delta}_\mathcal{B} A_k = - \frac{1}{T} \left( \mathbb{E}_k - T\partial_k \frac{\mu}{T} \right) = 0 \rightarrow \tilde{\delta}'_\mathcal{B} A_k \coloneqq - \frac{1}{T} \left( \mathbb{E}_k - T\partial_k \frac{\mu}{T} - \Gamma P_k \right) = 0,
\end{equation}
where $k$ runs only over the spatial indices. We can write this expression in a coordinate covariant manner as
	\begin{eqnarray}
		\delta'_{\mathcal{B}} A_{\mu} = \delta_{\mathcal{B}} A_{\mu} + \frac{1}{T} h_{\mu \nu} h^{\nu \rho} \Gamma_{\rho} \; .
	\end{eqnarray}
Subsequently the divergence of the canonical entropy current, in terms of $\delta'_{\mathcal{B}} A_{\mu}$, is given by
\begin{eqnarray}
		\label{Eq:FirstOrderPartEntropy}
		&\;& e^{-1} \partial_{\mu} \left( e S_{\mathrm{can}}^{\mu} \right) + \left( \beta^{\rho} - \frac{1}{T} \left( J^{\nu} - J^{\nu}_{(0)} \right) h_{\nu \sigma} h^{\sigma \rho} \right) \Gamma_\rho \nonumber \\
		&=&  \, \left(T^{\mu} - T^{\mu}_{(0)} \right) \mathcal{L}_\beta \tau_\mu - \frac{1}{2} \left(T^{\mu \nu} - T^{\mu \nu}_{(0)} \right) \mathcal{L}_\beta h_{\mu \nu} - \left( J^{\mu} - J^{\mu}_{(0)} \right) \delta'_\mathcal{B} A_\mu \; .
\end{eqnarray}
For the ideal fluid, where $J^{\mu} \rightarrow J^{\mu}_{(0)}$, the second term above is nothing more than \eqref{relaxation relation} with an order one electric field. To proceed further we need to classify corrections to the constitutive relations into hydrostatic, non-hydrostatic non-dissipative and dissipative pieces.}

\subsection{Hydrostatic terms}

{\noindent A useful decomposition of the types of terms one can encounter on the RHS of \eqref{Eq:FirstOrderPartEntropy} is given by splitting each of the pieces involving the constitutive relations into hydrostatic (HS), non-hydrostatic non-dissipative (NHS) and dissipative (D) corrections \cite{deBoer:2020xlc,Haehl:2014zda,Haehl:2015pja} i.e.:
\begin{equation}
	\label{Eq:CorrectionDecomposition}
	\begin{split}
		T^{\mu} - T^{\mu}_{(0)} & = T^{\mu}_{\text{HS}} + T^{\mu}_{\text{NHS}} + T^{\mu}_{\text{D}} , \\
		T^{\mu \nu} - T^{\mu \nu}_{(0)} & = T^{\mu \nu}_{\text{HS}} + T^{\mu \nu}_{\text{NHS}} + T^{\mu \nu}_{\text{D}}, \\ 
		J^{\mu} - J^{\mu}_{(0)} & = J^{\mu}_{\text{HS}} + J^{\mu}_{\text{NHS}} + J^{\mu}_{\text{D}} \; .
	\end{split}
\end{equation}
Let us proceed in order to define and deal with these terms. We note first that there are two ways that the equations of motion \eqref{eom current} and \eqref{eom SEM tensor} can be identically satisfied at hydrostaticity. Either, the derivative of the current, $T^{\mu}$, $T^{\mu \nu}$ and $J^{\mu}$, is proportional to a hydrostaticity condition, products of hydrostaticity conditions or their derivatives or - alternatively - the current itself is constructed from such terms. The hydrostatic contributions are of the former type while the non-hydrostatic and dissipative contributions are the latter. Consequently, the NHS and D contributions in \eqref{Eq:CorrectionDecomposition} vanish upon imposing hydrostaticity conditions while the HS do not. Nevertheless the HS contributions are constructed such that upon substitution into the equation of motion the resultant expression vanishes when we consider hydrostatic flows.}

{\ In the absence of the relaxation term $\Gamma_{\rho} \equiv 0$, the hydrostatic parts of the constitutive relations, corresponding to stationary flows of the system, can be obtained from the equilibrium generating functional \cite{Jensen:2012jh,Armas:2020mpr}. The generating functional to order one in derivatives takes the form
\begin{equation}
	W =\int d^{d+1} x \, e \left[ P\left(T,\mu,h_{\mu \nu} u^\mu u^\nu\right) + \sum_i F_i\left(T, \mu, h_{\mu \nu} u^\mu u^\nu\right) \tilde{s}_{(1)}^{(i)} + \mathcal{O}(\partial^2) \right],
\end{equation}
with the scalar basis
\begin{equation}
	\begin{aligned}
		\tilde{s}_{(1)}= & \left\{ \nu^\mu \partial_\mu T,  \nu^\mu \partial_\mu \mu,  \nu^\mu \partial_\mu (h_{\rho \sigma} u^\rho u^\sigma) \right\} \; .
	\end{aligned}
\end{equation}
Varying with respect to the background fields \eqref{Eq:ChargeDefs}, we find the following HS constitutive relations at order one in derivatives
\begin{subequations}
	\label{Eq:HSconstitutiverelations}
\begin{eqnarray}
J_{\mathrm{HS}}^\mu &= & n u^\mu + \frac{\partial F_0}{\partial \mu} u^\mu \nu^\rho \partial_\rho T + \frac{\partial F_1}{\partial \mu} u^\mu \nu^\rho \partial_\rho \mu + \frac{\partial F_2}{\partial \mu} u^\mu \nu^\nu \partial_\nu \left( h_{\rho \sigma} u^\rho u^\sigma \right) \nonumber \\
&\;&- \frac{1}{e} \partial_\nu \left( e \nu^\nu F_1 \right) u^\mu  +  \mathcal{O}\left(\partial^2\right) \; ,  \\
T_{\mathrm{HS}}^\mu &=& \epsilon u^\mu + \left( \frac{\partial F_0}{\partial T} T u^{\mu} + \frac{\partial F_0}{\partial \mu} \mu u^{\mu} + 2 \frac{\partial F_0}{\partial u^2} h_{\rho \sigma} u^{\rho} u^{\sigma} u^{\mu} \right) \nu^\lambda \partial_\lambda T \nonumber \\
&\;& + \left( \frac{\partial F_1}{\partial T} T u^{\mu} + \frac{\partial F_1}{\partial \mu} \mu u^{\mu} + 2 \frac{\partial F_1}{\partial u^2} h_{\rho \sigma} u^{\rho} u^{\sigma} u^{\mu} \right) \nu^\lambda \partial_\lambda \mu \nonumber \\
&\;& + \left( \frac{\partial F_2}{\partial T} T u^{\mu} + \frac{\partial F_2}{\partial \mu} \mu u^{\mu} + 2 \frac{\partial F_2}{\partial u^2} h_{\rho \sigma} u^{\rho} u^{\sigma} u^{\mu} \right) \nu^\lambda \partial_\lambda \left( h_{\alpha \beta} u^{\alpha} u^{\beta} \right) \nonumber \\
&\;& - \frac{1}{e} \partial_\rho \left( e F_0 \nu^\rho \right) T u^{\mu} - \frac{1}{e} \partial_\rho \left( e F_1 \nu^\rho \right) \mu u^{\mu} - \frac{2}{e} \partial_\rho \left( e F_2 \nu^\rho \right) h_{\alpha \beta} u^{\alpha} u^{\beta} u^{\mu} \nonumber \\ 
&\;&  + \mathcal{O}\left(\partial^2\right) \; , \\
T_{\mathrm{HS}}^{\mu \nu} &=& P h^{\mu \nu} + \rho_m u^\mu u^\nu + \left( h^{\mu \nu} F_0 \nu^\rho + 2 \frac{\partial F_0}{\partial u^2} u^{\mu} u^{\nu} \nu^{\rho} - 2 F_0 h^{\rho ( \mu} \nu^{\nu)} \right) \partial_\rho T \nonumber \\ 
&\;& + \left( h^{\mu \nu} F_1 \nu^\rho + 2 \frac{\partial F_1}{\partial u^2} u^{\mu} u^{\nu} \nu^{\rho} - 2 F_1 h^{\rho ( \mu} \nu^{\nu)} \right) \partial_\rho \mu \nonumber \\ 
&\;& + \left( h^{\mu \nu} F_2 \nu^\rho + 2 \frac{\partial F_2}{\partial u^2} u^{\mu} u^{\nu} \nu^{\rho} - 2 F_2 h^{\rho ( \mu} \nu^{\nu)} \right) \partial_\rho \left( h_{\alpha \beta} u^{\alpha} u^{\beta} \right) \nonumber \\ 
&\;& - \frac{2}{e} \partial_\rho \left( e F_2 \nu^{\rho} \right) u^{\mu} u^{\nu} + \mathcal{O}\left(\partial^2\right)  \;.
\end{eqnarray}
	\end{subequations}
In the absence of relaxation it is straightforward but tedious to verify that once these constitutive relations are substituted into the equation of motion \eqref{eom current} and \eqref{eom SEM tensor}, the result vanishes upon imposition of the hydrostaticity conditions \eqref{hydrostatic constraints}.}

{\ In the presence of relaxation, one cannot obtain the hydrostatic contributions from equilibrium generating functional. In particular, we must now satisfy
	\begin{eqnarray}
		\label{Eq:HydrostaticEOM}
		\partial_{\mu} T\indices{_{\mathrm{HS}}^\mu_\nu} - F_{\nu \mu} J^{\mu}_{\mathrm{HS}} - \Gamma_{\nu}^{\mathrm{HS}} = 0 \; , \qquad
		\partial_{\mu} J_{\mathrm{HS}}^{\mu} = 0 \; 
	\end{eqnarray}
upon imposition of the modified hydrostaticity condition \eqref{Eq:FinalHydrostaticityConstraint}. The result at $\mathcal{O}(\partial^{0})$ in the constitutive relations, or $\mathcal{O}(\partial)$ in the equations of motion, we already know. Namely
	\begin{eqnarray}
		\Gamma_{(1),\nu}^{HS} = \rho_{\mathrm{m}} \Gamma
		\begin{pmatrix}
			\vec{v}^2\\
			\vec{v}_i
		\end{pmatrix} \; .
	\end{eqnarray}
So we now consider $\mathcal{O}(\partial)$ in the constitutive relations, or $\mathcal{O}(\partial^2)$ in the equations of motion. With some work, it is possible to show from \eqref{Eq:HSconstitutiverelations}, that the order two piece of \eqref{Eq:HydrostaticEOM} vanishes using only the conditions presented in \eqref{conditions hydrostatic equilibrium 1}. Consequently, we are forced to take $\Gamma_{(2),\nu}^{\mathrm{HS}} \equiv 0$.}

{\ Now that we have the HS constitutive relations, we can turn to considering production of entropy once more. Again, in the absence of relaxation, we see that the divergence of the canonical entropy current \eqref{Eq:FirstOrderPartEntropy} vanishes upon imposing hydrostaticity. Outside of hydrostaticity it is not difficult to convince oneself that the hydrostatic terms do not have fixed sign\footnote{It is useful to imagine a hypothetical fluid where all transport coefficients are zero leaving only the hydrostatic contributions to the constitutive relations. Outside of hydrostaticity, entropy production should still be positive definite for this imaginary fluid.}. Thus, if we were to impose that the canonical entropy current must be positive definite, we would be forced to eliminate stationary configurations from our fluid through setting the relevant term in the equilibrium generating functional (i.e. $F_{i}$ or its thermodynamic derivative) to zero. To avoid eliminating swathes of stationary flows, we search for an $S^{\mu}_{\mathrm{non}}$ such that
\begin{equation}
	\label{Eq:NonCanonical}
	\begin{split}
		e^{-1} \partial_{\mu} \left( e S^{\mu}_{\text{non}} \right) = & - T^{\mu}_{\text{HS}} \mathcal{L}_\beta \tau_\mu + \frac{1}{2} T^{\mu \nu}_{\text{HS}} \mathcal{L}_\beta h_{\mu \nu} + J^{\mu}_{\text{HS}} \delta_{\mathcal{B}} A_\mu \; .
	\end{split}
\end{equation}
We can then rewrite the equation for the divergence of the canonical entropy current as
	\begin{subequations}
	\begin{eqnarray}
		 e^{-1} \partial_{\mu} \left( e S^{\mu} \right)
	 &=& \left( T_{\mathrm{NHS}}^{\mu} + T_{\mathrm{D}}^{\mu} \right) \mathcal{L}_\beta \tau_\mu - \frac{1}{2} \left(T_{\mathrm{NHS}}^{\mu \nu} + T_{\mathrm{D}}^{\mu \nu} \right) \mathcal{L}_\beta h_{\mu \nu} \nonumber \\
	 &\;& - \left( J_{\mathrm{NHS}}^{\mu} + J^{\mu}_{\mathrm{D}} \right) \delta_\mathcal{B} A_\mu \; , \\
	 S^{\mu} &=& S^{\mu}_{\mathrm{can}} + S^{\mu}_{\mathrm{non}} \; ,
	\end{eqnarray}
	\end{subequations}
where we remind the reader that this is the $\Gamma_{\rho}=0$ case. At order one $S_{\mathrm{non}}$ takes the form
	\begin{eqnarray}
		S_{\mathrm{non}}^{\mu} &=& \frac{\nu^\mu}{T}\left(F_0 \vec{u}^\lambda \partial_\lambda \mu+F_1 \vec{u}^\lambda \partial_\lambda T+F_2 \vec{u}^\lambda \partial_\lambda \vec{u}^2\right) \nonumber \\
&\;& -\frac{\vec{u}^\mu}{T}\left(F_0 \nu^\mu \partial_\mu \mu+F_1 \nu^\mu \partial_\mu T+F_2 \nu^\mu \partial_\mu \vec{u}^2\right) \; .
	\end{eqnarray}
It is upon the new entropy current $S^{\mu}$ that we subsequently impose positivity of entropy production as a constraint on our system. In doing so we have ensured all stationarity compatible terms in the constitutive relations are consistent with positivity of entropy production even outside of the hydrostatic limit. Said another way, hydrostatic terms do not contribute to entropy production.}

{\ In the presence of relaxation, $\Gamma_{\rho} \neq 0$, the hydrostatic contributions once again do not make positive definite contributions to the divergence of the canonical entropy current. The major difference to the case without relaxation is that we have more freedom to include these configurations. In particular, because we can fix the energy relaxation to mop up any inconvenient terms, we search for a non-canonical entropy current contribution $S^{\mu}_{\mathrm{non}}$ and a relaxation scalar $\Gamma^{\mathrm{non}}$ defined by
\begin{equation}
	\label{Eq:ModifiedNonCanonical}
	\begin{split}
		e^{-1} \partial_{\mu} \left( e S^{\mu}_{\text{non}} \right) + \Gamma^{\mathrm{non}} = & - T^{\mu}_{\text{HS}} \mathcal{L}_\beta \tau_\mu + \frac{1}{2} T^{\mu \nu}_{\text{HS}} \mathcal{L}_\beta h_{\mu \nu} + J^{\mu}_{\text{HS}} \delta_{\mathcal{B}}' A_\mu \; ,
	\end{split}
\end{equation}
such that
	\begin{subequations}
	\label{Eq:EntropyProductionNoHS}
	\begin{eqnarray}
		 e^{-1} \partial_{\mu} \left( e S^{\mu} \right)
	 &=& \left( T_{\mathrm{NHS}}^{\mu} + T_{\mathrm{D}}^{\mu} \right) \mathcal{L}_\beta \tau_\mu - \frac{1}{2} \left(T_{\mathrm{NHS}}^{\mu \nu} + T_{\mathrm{D}}^{\mu \nu} \right) \mathcal{L}_\beta h_{\mu \nu} - \left( J_{\mathrm{NHS}}^{\mu} + J^{\mu}_{\mathrm{D}} \right) \delta_\mathcal{B}' A_\mu \nonumber \\
	 &\;&  - \left( \beta^{\rho} - \frac{1}{T} \left( J_{\mathrm{NHS}}^{\nu} + J_{\mathrm{D}}^{\nu} \right) h_{\nu \sigma} h^{\sigma \rho} \right) \Gamma_\rho \geq 0 \; , \\
	 S^{\mu} &=& S^{\mu}_{\mathrm{can}} + S^{\mu}_{\mathrm{non}} \; , \qquad \Gamma^{\mathrm{non}}= \frac{1}{T} J^{\mu}_{\text{HS}} h_{\mu \sigma} h^{\sigma \rho} \Gamma_{\rho} \; .
	\end{eqnarray}
	\end{subequations}
We see that \eqref{Eq:ModifiedNonCanonical} is nothing more than a rewriting of \eqref{Eq:NonCanonical} which gives us the definitive identification of $\Gamma^{\mathrm{non}}$. The difference between the two scalar terms (divergence of the non-canonical entropy current and relaxation scalar) is that the latter must vanish in the no relaxation limit and thus is proportional to $\Gamma_{\rho}$. This requirement prevents one from taking terms in $e^{-1} \partial_{\mu} (e S^{\mu}_{\mathrm{non}})$ and hiding them in $\Gamma^{\mathrm{non}}$. With that said we now have eliminated all hydrostatic terms from our positivity of entropy production constraint \eqref{Eq:EntropyProductionNoHS} which once more makes all stationary configurations consistent with positivity of entropy production, as it did for the case without relaxation.}

\subsection{Non-hydrostatic, non-dissipative terms}

{\noindent Having eliminated problems in the hydrostatic sector, we can now consider the NHS and D contributions to \eqref{Eq:EntropyProductionNoHS}. The former terms are separated from the latter by being those that make no contribution to entropy production. In other words
	\begin{eqnarray}
	\label{Eq:NHSconstraint}
	 T^{\mu}_{\text{NHS}} \mathcal{L}_\beta \tau_\mu  - T^{\mu \nu}_{\text{NHS}} \frac{1}{2} \mathcal{L}_\beta h_{\mu \nu}  - J^{\mu}_{\text{NHS}} \delta'_\mathcal{B} A_\mu \equiv 0 \; .
	\end{eqnarray}
As we noted above, the non-hydrostatic non-dissipative part of the constitutive relations must be expressed in terms of the hydrostaticity constraints, their products and/or derivatives. So, at order one, the NHS constitutive relations must be linear combinations of
	\begin{eqnarray}
		\label{Eq:Relaxedhydrostaticconstraints}
		\mathcal{L}_{\beta} \tau_{\mu}  \; , \qquad \mathcal{L}_{\beta} h_{\mu \nu}  \; , \qquad
		\delta'_{\mathcal{B}} A_{\mu} \;.
	\end{eqnarray}
Consequently, \eqref{Eq:NHSconstraint} is a quadratic form in these hydrostaticity constraints. For the resultant quadratic form to fail to contribute to entropy production independently of the particular flow, the corresponding quadratic form matrix must be antisymmetric. Thus the non-hydrostatic non-dissipative contributions at first order in derivatives must take the form\footnote{Brackets on indices $T^{(ab)}$ indicate symmetrisation i.e. $T^{(ab)} = \frac{1}{2} \left( T^{ab} + T^{ba} \right)$.}
\begin{equation} \label{NHS part}
	\begin{split}
		\left(\begin{array}{c}
			T^{\mu}_{(1),\text{NHS}} \\
			T^{\mu \nu}_{(1), \text{NHS}} \\
			J^{\mu}_{(1), \text{NHS}}
		\end{array}\right) = 
		\left(\begin{array}{ccc}
			0 & N_{2}^{\mu (\rho \sigma)} & N_{1}^{\mu \rho} \\
			-N_{2}^{\rho (\mu \nu)} & 0 & N_{3}^{\rho (\mu \nu)} \\
			-N_{1}^{\rho \mu} & - N_{3}^{\mu (\rho \sigma)} & 0 \\
		\end{array}\right)
		\left(\begin{array}{c}
			\mathcal{L}_\beta \tau_\rho \\
			- \frac{1}{2} \mathcal{L}_\beta h_{\rho \sigma} \\
			- \delta'_{\mathcal{B}} A_\rho
		\end{array}\right),
	\end{split}
\end{equation}
so that
	\begin{align}
		& \hphantom{=} T^{\mu}_{(1),\text{NHS}} \mathcal{L}_\beta \tau_\mu  - T^{\mu \nu}_{(1),\text{NHS}} \frac{1}{2} \mathcal{L}_\beta h_{\mu \nu}  - J^{\mu}_{(1),\text{NHS}} \delta'_\mathcal{B} A_\mu \nonumber \\
		& = \left(\begin{array}{c}
			\mathcal{L}_\beta \tau_\rho \\
			- \frac{1}{2} \mathcal{L}_\beta h_{\rho \sigma}\\
			- \delta_{\mathcal{B}} A_\rho
		\end{array}\right)^{\mathrm{T}}
		\left(\begin{array}{ccc}
			0 & N_{2}^{\rho (\mu \nu)} & N_{1}^{\rho \mu} \\
			-N_{2}^{\mu (\rho \sigma)} & 0 & N_{3}^{\mu (\rho \sigma)} \\
			-N_{1}^{\mu \rho} & - N_{3}^{\rho (\mu \nu)} & 0 \\
		\end{array}\right)
		\left(\begin{array}{c}
			\mathcal{L}_\beta \tau_\mu \\
			- \frac{1}{2} \mathcal{L}_\beta h_{\mu \nu}\\
			- \delta'_{\mathcal{B}} A_\mu
		\end{array}\right) \equiv 0 \; .
	\end{align}
We note that the tensor elements of \eqref{NHS part} are order zero in derivatives.}

{\ Breaking boost symmetry by introducing a fixed background velocity $\vec{v}$ means that the $SO(d)$ spatial rotation symmetry reduces to the  $SO(d-1)$ subgroup. The tensors appearing in the coefficient matrix have to respect this reduced symmetry which preserves the absolute values of the velocity, which labels different states \cite{deBoer:2020xlc,Armas:2020mpr}. With this being said, the tensor elements of \eqref{NHS part} can be constructed using the $SO(d - 1)$ invariant tensors
\begin{equation}
	\left\{ \nu^\mu, \quad P^{\mu \nu}=h^{\mu \nu}-n^\mu n^\nu, \quad n^\mu=\frac{h^{\mu \nu} h_{\nu \rho} u^\rho}{\sqrt{u^2}} \right\}
\end{equation}
where $n^{\mu}$ denotes the normalised velocity with $u^2 = h_{\mu \nu} u^\mu u^\nu$ while $P^{\mu \nu}$ is a projector orthogonal to $n^{\mu}$. The most general $SO(d-1)$ symmetric rank-2 tensor we can construct from the reduced set that is order zero is
\begin{equation} \label{2-tensor}
	N^{\mu \nu}_1  \coloneqq \alpha_1 P^{\mu \nu} + \alpha_2 n^\mu n^\nu + 2 \alpha_3 n^{(\mu} \nu^{\nu)} + \alpha_4 \nu^\mu \nu^\nu,
\end{equation}
while the most general three-tensor, which is symmetric in two of the indices, is given by
\begin{equation} \label{3-tensor}
	\begin{split}
		N^{\rho (\mu \nu)}_p \coloneqq & \beta_{1,p} \nu^\rho P^{\mu \nu} +\beta_{2,p} n^{\rho} P^{\mu \nu} + 2 \beta_{3,p} P^{\rho(\mu}\nu^{\nu)} + 2 \beta_{4,p} P^{\rho (\mu} n^{\nu)} \\ 
		& + \beta_{5,p} n^\rho n^\mu n^\nu + \beta_{6,p} \nu^\rho n^\mu n^\nu + 2 \beta_{7,p} n^\rho n^{(\mu} \nu^{\nu )} + 2 \beta_{8,p} \nu^{\rho} n^{(\mu} \nu^{\nu)} \\
		& + \beta_{9,p} n^{\rho} \nu^{\mu} \nu^{\nu} + \beta_{10,p} \nu^{\rho} \nu^{\mu} \nu^{\nu}, 
	\end{split}
\end{equation}
with $p  = {2,3}$. We have obtained the most general tensor structures consistent with our symmetries and defined the 24 non-hydrostatic, non-dissipative transport coefficients. This completes our analysis of the NHS sector.}

\subsection{Dissipative terms}

{\noindent The dissipative terms lead to a non-zero production of entropy. Analogously to the NHS case, because at order one the constitutive relations must be linear combinations of the hydrostaticity constraints \eqref{Eq:Relaxedhydrostaticconstraints}, the dissipative contributions can be written as the following quadratic form
\begin{equation}
	\begin{split}
		0 \leq & \quad T^{\mu}_{(1), \text{D}} \mathcal{L}_\beta \tau_\mu - T^{\mu \nu}_{(1),\text{D}} \frac{1}{2} \mathcal{L}_\beta h_{\mu \nu} - J^{\mu}_{(1), \text{D}} \delta'_\mathcal{B} A_\mu \\
		= & \left(\begin{array}{c}
			\mathcal{L}_\beta \tau_\rho \\
			- \frac{1}{2} \mathcal{L}_\beta h_{\rho \sigma}  \\
			- \delta_{\mathcal{B}} A_\rho
		\end{array}\right)^{\mathrm{T}}
		\left(\begin{array}{ccc}
			D_{1}^{\rho \mu} & D_{2}^{\rho (\mu \nu)} & D_{3}^{\rho \mu} \\
			D_{2}^{\mu (\rho \sigma)} & D_{4}^{(\rho \sigma) (\mu \nu)} & D_{5}^{\mu (\rho \sigma)} \\
			D_{3}^{\mu \rho} & D_{5}^{\rho (\mu \nu)}  & D_{6}^{\rho \mu} \\
		\end{array}\right)
		\left(\begin{array}{c}
			\mathcal{L}_\beta \tau_\mu \\
			- \frac{1}{2} \mathcal{L}_\beta h_{\mu \nu} \\
			- \delta'_{\mathcal{B}} A_\mu
		\end{array}\right),
	\end{split}
\end{equation}
where the symmetric coefficient matrix, in contrast to the antisymmetric matrix of \eqref{NHS part}, allows for entropy production.  Its contribution to the constitutive relations is simply given by
\begin{equation}
	\begin{split}
		\left(\begin{array}{c}
			T^{\mu}_{(1),\text{D}} \\
			T^{\mu \nu}_{(1), \text{D}} \\
			J^{\mu}_{(1), \text{D}}
		\end{array}\right) = 
		\left(\begin{array}{ccc}
			D_{1}^{\mu \rho} & D_{2}^{\mu (\rho \sigma)} & D_{3}^{\mu \rho} \\
			D_{2}^{\rho (\mu \nu)} & D_{4}^{(\mu \nu) (\rho \sigma)} & D_{5}^{\rho (\mu \nu)} \\
			D_{3}^{\rho \mu} & D_{5}^{\mu (\rho \sigma)} & D_{6}^{\mu \rho} \\
		\end{array}\right)
		\left(\begin{array}{c}
			\mathcal{L}_\beta \tau_\rho \\
			- \frac{1}{2} \mathcal{L}_\beta h_{\rho \sigma}  \\
			- \delta'_{\mathcal{B}} A_\rho
		\end{array}\right).
	\end{split}
\end{equation}
The rank-2 and -3 tensors $D_{p}^{\mu \nu}$ and $D_{q}^{\mu (\nu \rho)}$ of this matrix take the same form as (\ref{2-tensor}) and (\ref{3-tensor}) with appropriate renaming of transport coefficients, while the rank-4 tensor is given by
\begin{equation} \label{rank 4 tensor}
	\begin{aligned}
		D_{4}^{(\mu \nu) (\rho \sigma)} = & \, \mathfrak{t}\left(P^{\mu \rho} P^{\nu \sigma}+P^{\mu \sigma} P^{\nu \rho}-\frac{2}{d-1} P^{\mu \nu} P^{\rho \sigma}\right)+ 4 \gamma_1 \nu^{(\mu} n^{\nu)} n^{(\rho} \nu^{\sigma)} \\
		& +\gamma_2 n^\mu n^\nu n^\rho n^\sigma+\gamma_3 P^{\mu \nu} P^{\rho \sigma} + 4 \gamma_4 \nu^{(\mu} P^{\nu)(\rho} \nu^{\sigma)} \\ 
		& +\gamma_5 \left(P^{\mu \rho} n^\nu n^\sigma+P^{\nu \rho} n^\mu n^\sigma+P^{\mu \sigma} n^\nu n^\rho+P^{\nu \sigma} n^\mu n^\rho\right) \\
		& +\gamma_6 \left(P^{\rho \sigma} n^\mu n^\nu+P^{\mu \nu} n^\rho n^\sigma\right) - 4 \gamma_7 \left(\nu^{(\mu} P^{\nu)(\rho} n^{\sigma)}+n^{(\mu} P^{\nu)(\rho} \nu^{\sigma)}\right) \\
		& - 2 \gamma_8 \left(\nu^{(\mu} n^{\nu)} P^{\rho \sigma}+P^{\mu \nu} n^{(\rho} \nu^{\sigma)}\right) \\
		& - 2 \gamma_9 \left(\nu^{(\mu} n^{\nu)} n^\rho n^\sigma+n^\mu n^\nu n^{(\rho} \nu^{\sigma)}\right).
	\end{aligned}
\end{equation}
We have now obtained the most general tensor structures consistent with our symmetries and defined the 42 dissipative transport coefficient terms.}

{\ Finally, we can turn to the constraint on the relaxation terms given by the final term in \eqref{Eq:EntropyProductionNoHS}. Their contribution to entropy production must vanish as the relevant term is not positive definite. We see therefore that the energy relaxation term in our construction must receive derivative corrections which at first order in derivatives in the FSCC limit are given by \cite{Martinoia:2024cbw}
\begin{equation} \label{energyrelaxation}
	\hat\Gamma_\varepsilon = \rho_{m}  \Gamma v_{j} \left( nv^{j} + J^{j}_{(1),\text{NHS}} + J^{j}_{(1), \text{D}} \right) + \mathcal{O}(\partial^3)  \; .
\end{equation}
This completes our first aim of analysing relaxation at order one in derivatives for a modified hydrostaticity constraint of the form \eqref{Eq:FinalHydrostaticityConstraint}.}

\subsection{Moving to the thermodynamic density frame}

{\noindent Up till now we have used the ``thermodynamic frame'' where the thermal vector $\beta^\mu$ and the $U(1)$ gauge parameter $\Lambda$ are chosen to assume the ideal order values, i.e. such that they do not receive derivative corrections given by
\begin{equation}
\beta^\mu = \frac{u^\mu}{T}, \quad \quad \Lambda = \frac{\mu - u^\mu A_\mu}{T}.
\end{equation}
However, as has been pointed out in \cite{Armas:2020mpr}, this choice does not constitute a complete fixing in the sense that certain redefinitions of the hydrodynamic variables $u^\mu,T$ and $\mu$ leave the dynamics of the theory unaltered.}

{\ To alleviate the frame ambiguity, we first note that the equations of motion can be expressed as a linear combination of $\mathcal{L}_\beta \tau_\mu$, $\mathcal{L}_\beta h_{\mu \nu}$ and $\delta_{\mathcal{B}} A_\mu$. Correspondingly they can be used to eliminate any $(d+1+2)$ number of linear combinations of these terms in higher order constitutive relations, where $\text{d}$ denotes the number of spatial dimensions. This results in $d(d+5)/2$ independent components. To fix the remaining freedom we choose the ``thermodynamic density frame'' \cite{Armas:2020mpr} for which the independent components that the constitutive relations are being expanded in are  $h^{\mu \nu} \mathcal{L}_{\beta} \tau_{\nu}$, $h^{\mu \rho} h^{\nu \sigma} \mathcal{L}_{\beta} h_{\rho \sigma}$ and $h^{\mu \nu} \delta_{\beta} A_{\nu}$. The non-hydrostatic sector of this frame coincides with the one of the "density frame" which is defined such that the energy, momentum and particle number are equal to their ideal order values
\begin{equation}
T^\mu \tau_\mu = \epsilon, \quad \quad T^{0 \mu} = \rho_m \left( u^\mu - \nu^\mu \right), \quad \quad J^\mu \tau_\mu = n.
\end{equation}
}

{\ Then,  only the spatial components of the tensor structures of the coefficient matrix occur and the constitutive relations reduce to \cite{Armas:2020mpr}
\begin{equation}
	\begin{split}
		\left(\begin{array}{c}
			T^{\mu}_{(1),\text{NHS}} \\
			T^{\mu \nu}_{(1), \text{NHS}} \\
			J^{\mu}_{(1), \text{NHS}}
		\end{array}\right) = T
		\left(\begin{array}{ccc}
			0 & N_{2}^{\mu (\rho \sigma)} & N_{1}^{\mu \rho} \\
			-N_{2}^{\rho (\mu \nu)} & 0 & N_{3}^{\rho (\mu \nu)} \\
			-N_{1}^{\rho \mu} & - N_{3}^{\mu (\rho \sigma)} & 0 \\
		\end{array}\right)
		\left(\begin{array}{c}
			\mathcal{L}_\beta \tau_\rho \\
			- \frac{1}{2} \mathcal{L}_\beta h_{\rho \sigma} \\
			- \delta_{\mathcal{B}} A_\rho
		\end{array}\right),
	\end{split}
\end{equation}
and
\begin{equation}
	\begin{split}
		\left(\begin{array}{c}
			T^{\mu}_{(1),\text{D}} \\
			T^{\mu \nu}_{(1), \text{D}} \\
			J^{\mu}_{(1), \text{D}}
		\end{array}\right) =
		T \left(\begin{array}{ccc}
			D_{1}^{\mu \rho} & D_{2}^{\mu (\rho \sigma)} & D_{3}^{\mu \rho} \\
			D_{2}^{\rho (\mu \nu)} & D_{4}^{(\mu \nu) (\rho \sigma)} & D_{5}^{\rho (\mu \nu)} \\
			D_{3}^{\rho \mu} & D_{5}^{\mu (\rho \sigma)} & D_{6}^{\mu \rho} \\
		\end{array}\right)
		\left(\begin{array}{c}
			\mathcal{L}_\beta \tau_\rho \\
			- \frac{1}{2} \mathcal{L}_\beta h_{\rho \sigma}  \\
			- \delta_{\mathcal{B}} A_\rho
		\end{array}\right).
	\end{split}
\end{equation}
with
\begin{equation} \label{tensor structures coefficient matrix}
	\begin{split}
		N^{\mu \nu}_1 & \coloneqq \alpha_{1} P^{\mu \nu} + \alpha_{2} \vec{n}^\mu \vec{n}^\nu \\
		N^{\mu (\rho \sigma)}_p & \coloneqq \beta_{p,2} \vec{n}^{\mu} P^{\rho \sigma} + 2 \beta_{p,4} P^{\mu (\rho} \vec{n}^{\sigma)} + \beta_{p,5} \vec{n}^\mu \vec{n}^\rho \vec{n}^\sigma \\
		D^{\mu \nu}_{q} & \coloneqq \bar{\alpha}_{q,1} P^{\mu \nu} + \bar{\alpha}_{q,2} \vec{n}^\mu \vec{n}^\nu \\
		D^{\mu (\rho \sigma)}_r & \coloneqq \bar{\beta}_{r,2} \vec{n}^{\mu} P^{\rho \sigma} + 2 \bar{\beta}_{r,4} P^{\mu (\rho} \vec{n}^{\sigma)} + \bar{\beta}_{r,5} \vec{n}^\mu \vec{n}^\rho \vec{n}^\sigma \\
		D_{4}^{(\mu \nu) (\rho \sigma)} & \coloneqq \mathfrak{t}\left(P^{\mu \rho} P^{\nu \sigma}+P^{\mu \sigma} P^{\nu \rho}-\frac{2}{d-1} P^{\mu \nu} P^{\rho \sigma}\right) + \gamma_2 \vec{n}^\mu \vec{n}^\nu \vec{n}^\rho \vec{n}^\sigma +\gamma_3 P^{\mu \nu} P^{\rho \sigma} \\
		& \quad +4 \gamma_5 \vec{n}^{(\mu} P^{\nu) (\rho} \vec{n}^{\sigma)} +\gamma_6 \left( \vec{n}^\mu \vec{n}^\nu P^{\rho \sigma}+P^{\mu \nu} \vec{n}^\rho \vec{n}^\sigma\right),
	\end{split}
\end{equation}
with $p=\{2,3\}$, $q=\{1,3,6\}, \,r=\{2,5\}$, $\vec{n}^{\mu} = \left(u^\mu - \nu^\mu \right)/|u^\mu - \nu^\mu|$ and $P^{\mu \nu} = h^{\mu \nu} - \vec{n}^\mu \vec{n}^\nu$ as before.}

\section{Conductivities}
\label{sec:conductivity}

{\noindent We now turn our interest to the analysis of the linear response of the such relaxed fluids, in particular we look at the AC conductivities. We define a response matrix of the following form
\begin{equation} \label{linear response}
	\left(\begin{array}{c}
		\delta J_i \\
		\delta Q_i \\
		\delta P_i
	\end{array}\right)=\left(\begin{array}{ccc}
		\sigma_{i j} & T \alpha_{i j} & \zeta_{ij}^1\\
		T \bar{\alpha}_{i j} & T \kappa_{i j} & \zeta_{ij}^2\\
		\zeta^3_{ij} & \zeta^4_{ij} & \zeta^5_{ij} \\
	\end{array}\right)\left(\begin{array}{c}
		\delta E_j \\
		\delta\left(-\frac{\partial_j T}{T} \right) \\
		\delta v_{0 j}
	\end{array}\right), 
\end{equation}
which specifies how each of the charge currents (electric $\delta J^i$,  heat $\delta Q^i= \delta J_E^{i} - \mu \delta J^i \equiv \delta T^i_{\ 0} - \mu \delta J^i$ and spatial momentum $P^{i}$) respond to perturbations of the electric field, temperature and spatial velocity. In contrast to the relativistic case, in the boost-agnostic case the electric current $J^i$ and the momentum current $P^i$ are independent. Consequently, we extend the standard conductivity matrix by the coefficients $\zeta^i$s to capture the full transport behaviour.}

{\ To obtain the matrix in \eqref{linear response} we linearise and solve the hydrodynamic equations in the presence of the sources indicated in the vector of \eqref{linear response}.  We consider small fluctuations of our fluid away from a stationary configuration with zero spatial velocity, constant temperature $T$ and chemical potential $\mu$ i.e.
\begin{subequations}
\begin{eqnarray}
		&\;& \mu(x^\mu) = \mu + \delta \mu(x^\mu), \qquad  T(x^\mu) = T + \delta T(x^\mu), \\
		&\;& u^\mu(x^\mu) = \left(1, \delta v_x(x^\mu), \delta v_y(x^\mu), \delta v_z(x^\mu) \right) \; .
\end{eqnarray}
\end{subequations}
Subsequently, the spatially Fourier-transformed linearised conservation equations assume the form
\begin{equation}
\partial_t \varphi_a(t,\bold{k}) + M_{ab}(\bold{k}) \varphi_b(t,\bold{k}) = 0,
\end{equation}
where $\varphi_a$ denotes the fluctuations of the hydrodynamic variables $\varphi_a = (\delta \epsilon, \delta P^i, \delta n)$. The matrix $M_{ab}$ depends on the specific expression for the  conservation equations and constitutive relations of above.  The retarded Green's function are then given by \cite{Kovtun:2012rj}
\begin{equation}
\label{Eq:MartinKadanoffGreen}
G^{R}_{ab}(z,\bold{k}) = - (1 + i z K^{-1}(z,\bold{k}))_{ac} \chi_{cb},
\end{equation}
with $K_{ab} = - i z \delta_{ab} + M_{ab}(\bold{k})$ and $\chi_{ab}$ the susceptibility matrix
\begin{equation}
\chi_{ab} = \frac{\partial \varphi_a}{\partial \lambda_b},
\end{equation}
where $\lambda_a = (\frac{\delta T}{T}, \delta v^i, \delta\mu-\frac{\mu}{T}\delta T )$.}

{\ The tensors (\ref{tensor structures coefficient matrix}) appearing in the constitutive relations have been expressed in terms of normalised velocities. In order for the corresponding terms to be regular for vanishing spatial velocities,  which characterises the state around which we fluctuate, we need to assume that the associated transport coefficients that multiply the tensor structures satisfy a Taylor series expansion in $u^2 \equiv |\vec{u}|^2$. For regularity, we then have to demand the transport coefficients to satisfy the conditions listed in appendix \ref{Appendix:lowvlimit}.}

{\ With this above caveat accepted, the AC conductivities given by the $\vec{k}\rightarrow \vec{0}$ limit of \eqref{Eq:MartinKadanoffGreen} from our hydrodynamic model are\footnote{We are setting to zero the hydrostatic terms, since they take the same form with or without relaxation rates.}
\begin{subequations}\label{conductivities2}
\begin{align} 
		\sigma(\omega,\vec{0}) & = \bar{\alpha}_{6,1}+\frac{n (n -\Gamma \rho_m \bar{\alpha}_{6,1})}{\rho_m (\Gamma -i \omega )}, \label{eqn:electric_conductivity}\\
		 \alpha(\omega,\vec{0}) & = \frac{\bar{\alpha}_{3,1} + \alpha_{1} -\bar{\alpha}_{6,1} \mu}{T}+\frac{s (n - \bar{\alpha}_{6,1} \Gamma \rho_m)}{\rho_m (\Gamma -i \omega )}, \\
		\bar{\alpha}(\omega,\vec{0}) & = \frac{ \bar{\alpha}_{3,1} -\alpha_{1} - \bar{\alpha}_{6,1} \mu}{T} +\frac{n  \left(s + \frac{\Gamma \rho_m (\bar{\alpha}_{6,1} \mu + \alpha_{1} -\bar{\alpha}_{3,1})}{T} \right)}{\rho_m (\Gamma -i \omega )}, \\
		\kappa(\omega,\vec{0}) & = \frac{\bar{\alpha}_{1,1} - 2 \bar{\alpha}_{3,1} \mu + \bar{\alpha}_{6,1} \mu^2}{T} +\frac{s(sT +\Gamma  \rho_m ( - \bar{\alpha}_{3,1} + \alpha_{1} + \bar{\alpha}_{6,1} \mu) )}{\rho_m (\Gamma -i \omega )}, \\
		\zeta^1(\omega,\vec{0}) & =	\frac{n - \bar{\alpha}_{6,1} \Gamma \rho_m}{\Gamma - i \omega}, \\
		\zeta^2(\omega,\vec{0}) & = \frac{s T + \Gamma \rho_m \left( \bar{\alpha}_{6,1} \mu + \alpha_{1} - \bar{\alpha}_{3,1} \right)}{{\Gamma - i \omega}}	, \\
	    \zeta^3(\omega,\vec{0}) & = \frac{n}{\Gamma - i \omega}, \\
		\zeta^4(\omega,\vec{0}) & = \frac{s T}{\Gamma - i \omega}, \\
		\zeta^5(\omega,\vec{0}) & = \frac{\rho_m}{\Gamma - i \omega} \; ,
\end{align}
\end{subequations}
where we have not imposed Onsager reciprocity (for example $\alpha \neq \bar{\alpha}$). One motivation for avoiding the imposition of this symmetry comes from our modified hydrostaticity condition \eqref{Eq:FinalHydrostaticityConstraint}, which we repeat here for ease:
	\begin{eqnarray}
		\mathbbm{E}_{i} - \partial_{i} \mu - \Gamma P_{i} &=& 0 \; .
	\end{eqnarray}
All other hydrostaticity conditions \eqref{conditions hydrostatic equilibrium 1} have definite eigenvalues under time reversal; \eqref{Eq:FinalHydrostaticityConstraint} is the exception. It explicitly breaks time reversal symmetry and is fundamental to defining the stationary states. However, in the case of vanishing relaxations and taking $\alpha_{1}=0$ (which breaks Onsager relations even in the absence of relaxations \cite{Armas:2020mpr}), this expression gains a definitive sign and indeed:
\begin{equation}
	\alpha = \bar{\alpha} = \frac{\bar{\alpha}_{3,1} - \bar{\alpha}_{6,1} \mu}{T} + \frac{i n s}{\rho_m \omega} \; , \qquad \zeta^1 = \zeta^3 \; , \qquad \zeta^2 = \zeta^4 \; .
\end{equation}
We leave a more detailed analysis of the consequences of Onsager reciprocity for the next section.}

The result obtained in equation \eqref{eqn:electric_conductivity} for the electric conductivity closely resembles certain expressions obtained using holographic methods. To make contact with the literature, call $\bar{\alpha}_{6,1}=\sigma_0$ the intrinsic conductivity which appears in relativistic hydrodynamics and further notice that $\sigma(\omega\rightarrow0)=\sigma_\text{DC}=n^2/\rho_m\Gamma$. With these definitions we can rewrite \eqref{eqn:electric_conductivity} as
\begin{equation}
	\sigma(\omega)=\sigma_0+\frac{\sigma_\text{DC}-\sigma_0}{1-i\omega\tau}
\end{equation}
where $\tau=\Gamma^{-1}$. This expression clearly differs from the standard hydrodynamic conductivity, which is the sum of a coherent Drude part and an incoherent term parametrized by $\sigma_0$, but appears in certain holographic models when the momentum-breaking parameter $\Gamma$ becomes large enough \cite{Zhou:2015qui,Davison:2015bea,Fu:2022jqn}.

{\ We complete our goal of obtaining the conductivities by identifying an incoherent current. In the case of relativistic hydrodynamics with a $U(1)$ charge, the incoherent current is defined by
	\begin{eqnarray}
		&\;& J^{i}_{\mathrm{inc}.} = \frac{s T}{\epsilon + p} J^{i} - \frac{n}{\epsilon+p} Q^{i} \; .
	\end{eqnarray}
This current has two nice properties
	\begin{equation}
		\langle J^{i}_{\mathrm{inc}} J^{j}_{\mathrm{inc}} \rangle(\omega,\vec{0}) = \bar{\alpha}_{6,1} \delta^{ij} := \sigma_{\mathrm{inc}} \delta^{ij} \; , \qquad
		\langle J^{i}_{\mathrm{inc}} P^{j} \rangle(\omega,\vec{0}) = 0 \; .
	\end{equation}
Consequently, it represents the charge not carried along by spatial momentum during a flow. In the boost-agnostic case, we seek to define a similar current, but now the momentum and heat are distinct - unlike the relativistic case. Taking the most generic linear combination of the spatial currents gives
	\begin{eqnarray}
		J^{i}_{\mathrm{inc}.} = c_{0} J^{i} - c_{1} Q^{i} - c_{2} P^{i} \; ,
	\end{eqnarray}
and we would like to identify the part of the charge current that is carried neither by heat, nor spatial momentum i.e. we impose the constraints
	\begin{eqnarray}
		\langle J^{i}_{\mathrm{inc}} J^{j}_{\mathrm{inc}} \rangle(\omega,\vec{0}) = \bar{\alpha}_{6,1} \delta^{ij} \; , \qquad  \langle J^{i}_{\mathrm{inc}} Q^{j} \rangle(\omega,\vec{0}) = 0 \; , \qquad \langle J^{i}_{\mathrm{inc}} P^{j} \rangle(\omega,\vec{0}) = 0 \; . \qquad
	\end{eqnarray}
Using the zero wave-vector, finite frequency expressions \eqref{conductivities2}, we soon find that the coefficients $c_{0}$, $c_{1}$ and $c_{2}$ are given in terms of transport terms by
	\begin{subequations}
	\label{Eq:IncCoeffs}
	\begin{eqnarray}
		c_0 &=& \pm \frac{\sqrt{\bar{\alpha}_{6,1}} \sqrt{\mu  (\mu  \bar{\alpha}_{6,1}-2 \bar{\alpha}_{3,1})+\bar{\alpha}_{1,1}}}{\sqrt{ \bar{\alpha}_{6,1} \bar{\alpha}_{1,1} - \bar{\alpha}_{3,1}^2 + \alpha_{1}^2 }} \\
		c_1 &=& \pm \frac{\sqrt{\bar{\alpha}_{6,1}} (-\mu  \bar{\alpha}_{6,1}+\bar{\alpha}_{3,1}+\alpha_{1})}{\sqrt{(\mu  (\mu  \bar{\alpha}_{6,1}-2 \bar{\alpha}_{3,1})+\bar{\alpha}_{1,1}) \left( \bar{\alpha}_{6,1} \bar{\alpha}_{1,1} - \bar{\alpha}_{3,1}^2 + \alpha_{1}^2 \right)}} \\
		c_2 &=& \pm \frac{\sqrt{\bar{\alpha}_{6,1}} (\rho  (\mu  (\alpha_{1}-\bar{\alpha}_{3,1})+\bar{\alpha}_{1,1})-(p+\epsilon ) (-\mu  \bar{\alpha}_{6,1}+\bar{\alpha}_{3,1}+\alpha_{1}))}{\rho_m \sqrt{(\mu  (\mu  \bar{\alpha}_{6,1}-2 \bar{\alpha}_{3,1})+\bar{\alpha}_{1,1}) \left( \bar{\alpha}_{6,1} \bar{\alpha}_{1,1} - \bar{\alpha}_{3,1}^2 + \alpha_{1}^2 \right)}}
	\end{eqnarray}
	\end{subequations}
If we try to recover the relativistic incoherent current from these expressions we will find that they diverge as we are overconstraining our system by requiring that $\langle J^{i}_{\mathrm{inc}.} Q^{i} \rangle(\omega,\vec{0}) =0$.}

{\ The expressions given above \eqref{Eq:IncCoeffs} do not satisfy what we would desire as the coefficients are given as functions of transport coefficients and not thermodynamic susceptibilities and/or global thermodynamic parameters of the system (e.g. $n$, $\epsilon$, $P$...). We invoke another property of the incoherent current to remedy this fault, namely, we impose
	\begin{equation}
		\langle J^{i}_{\mathrm{inc}} Q^{j} \rangle(0,\vec{k}) = 0 \; , \qquad
		\langle J^{i}_{\mathrm{inc}} P^{j} \rangle(0,\vec{k}) = 0 \; .
	\end{equation}
Using the above constraints, it is not hard to show that $c_{0}$ and $c_{1}$ can be expressed in terms of thermodynamic susceptibilties and the arbitrary coefficient $c_{2}$,
	\begin{subequations} \label{Eq:IncCoeffs2}
		\begin{eqnarray}
			c_0 &=& - c_{2} \left( \frac{\chi_{PP} \chi_{QQ}- \chi_{QP}^2}{\chi_{JQ} \chi_{QP}-\chi_{JP} \chi_{QQ}} \right) \; ,\\
			c_1 &=& - c_{2} \left( \frac{\chi_{JQ} \chi_{PP}- \chi_{JP} \chi_{QP}}{\chi_{JQ} \chi_{QP}-\chi_{JP} \chi_{QQ}} \right) \; ,
		\end{eqnarray}
	\end{subequations}
where $\chi_{ab}$ represents the thermodynamic susceptibility for the currents $a$ and $b$. Equating these expressions to \eqref{Eq:IncCoeffs} we find that it is sufficient to constrain $\alpha_{1}$ in terms of the other transport coefficients
	\begin{eqnarray}
		\alpha_{1} & =& \mu \bar{\alpha}_{6,1} - \bar{\alpha}_{3,1} + \left( \mu^2  \bar{\alpha}_{6,1}-2 \mu \bar{\alpha}_{3,1}+ \bar{\alpha}_{1,1} \right) \frac{\chi_{JQ} \chi_{PP}-\chi_{JP} \chi_{PQ}}{\chi_{PP} \chi_{QQ}-\chi_{QP}^2} \; ,
	\end{eqnarray}
and choose $c_{0}$ and $c_{1}$ as in \eqref{Eq:IncCoeffs2} to ensure that there is at least one incoherent current satisfying the desired properties. Of course, there is in fact a whole family among which we can choose a candidate.}

\section{The consequences of imposing Onsager reciprocity}
\label{sec:onsager}

{\noindent We now turn to consider the consequence of imposing microscopic time reversal covariance on the system. From the conductivities of the previous section \eqref{conductivities2}, it is easy to see that we must require that
\begin{equation} \label{conductivity constraint}
	\alpha_{1} =0 \; , \qquad \bar{\alpha}_{3,1} = \left(\frac{s T}{n} + \mu \right) \bar{\alpha}_{6,1} \; ,
\end{equation}
as $\alpha=\bar{\alpha}$ in a time reversal covariant system. However, this is not the only constraint imposed by time reversal covariance. In particular, it is necessary for all the correlators to satisfy
\begin{equation}
\label{Eq:Onsager}
G^{R}_{ab}(\omega, \bold{k}) = \eta_{a} \eta_{b} G^{R}_{ba} (\omega, -\bold{k}), 
\end{equation}
with $\eta_a = \pm 1$ being the time-reversal eigenvalue of the field $\varphi_a$. Applying this relation \eqref{Eq:Onsager} to the retarded Green's function of linear response theory \eqref{Eq:MartinKadanoffGreen}, leads to the following conditions that need to be satisfied
\begin{equation} \label{onsager}
\chi S M^T(-\bold{k}) - M(\bold{k}) \chi S = 0, \qquad S = \text{diag}(1,-1,-1,-1,1) \; ,
\end{equation}
for any theory with our content of current. The matrix $S$ is given by the time-reversal eigenvalues of the $\varphi$s.}

{\ From \eqref{onsager} we see that  in our model, if we want to respect microscopic time reversal symmetry in the effective correlators at non-zero frequency and wavevector for a state a zero velocity, we have to impose
\begin{equation}
\label{Eq:MMatrix}
\left(  \begin{array}{ccc}
 0 & - i \Gamma \rho_m q_i \bar{\alpha}_{3,1} &  0 \\
- i \Gamma \rho_m q_i \bar{\alpha}_{3,1} & 0 & -i \Gamma \rho_m q_i \bar{\alpha}_{6,1} \\
 0 & -i \Gamma \rho_m q_i \bar{\alpha}_{6,1}  & 0 \\
\end{array} \right) = 0 \; ,
\end{equation}
in addition to \eqref{conductivity constraint}. It is clear from the above matrix that in the standard boost agnostic case with vanishing relaxations, Onsager reciprocity is satisfied by \eqref{conductivity constraint} without having to impose any further constraints on the dissipative part of the constitutive relations.}

{\ In the relaxed case however, the transport coefficients $\bar{\alpha}_{6,1} $ and $\bar{\alpha}_{3,1}$ have to vanish to satisfy \eqref{Eq:MMatrix}.
Naturally this is consistent with the conductivity constraint (\ref{conductivity constraint}) as it trivialises the equation. Consequently the thermoelectric conductivities in (\ref{conductivities2}) reduce to
\begin{equation}
		\sigma = \frac{n^2}{\rho_m (\Gamma -i \omega )} \; , \qquad
		\alpha = \bar{\alpha} = \frac{n s}{\rho_m (\Gamma -i \omega )} \; , \qquad
		\kappa = \frac{s^2 T}{\rho_m (\Gamma -i \omega )} \; .
\end{equation}
Thus we have reached one of our major conclusions: if we impose positivity of entropy production and Onsager reciprocity, then we find that the incoherent conductivities must vanish and we obtain the Drude result for the electric conductivity
\begin{equation}
\sigma = \frac{\sigma_{\text{DC}}}{1 -i \omega \Gamma^{-1}}
\end{equation} 
with the DC conductivity being given by
\begin{equation}
\sigma_{\text{DC}} = \frac{n^2}{\rho_m \Gamma}.
\end{equation}
Thus, the incoherent conductivity can only appear if the system does not form a steady state or if we violate Onsager reciprocity. This result could be anticipated from the earlier work on stationary flows (where there is also a non-trivial flow of charge in the stationary state). However, we have now shown indeed that the DC conductivity obtained from the full dissipative theory and the stationary sector must agree. This conclusion is independent of the time-reversal symmetry properties of the retarded correlators, since even the conductivity in \eqref{eqn:electric_conductivity} has a coherent DC limit.} Contrariwise, in systems where relaxation rates do not allow for our driven steady states, one can still potentially find a non-zero incoherent conductivity and preserve Onsager relations \cite{Hartnoll:2007ih,Amoretti:2023vhe}.

\section{Discussion}
\label{sec:discussion}

{\noindent In this paper, we extended the discussion of \citep{Amoretti:2022ovc} on the feasibility of modifying hydrostaticity constraints using relaxation terms. Our analysis incorporates tensor structures that vanish at stationarity, including dissipative ones, enabling us to identify the first-order derivative corrections for energy relaxation (see equation \ref{energyrelaxation}). We then computed the thermo-electric conductivities of our system in the case where Onsager reciprocity is not imposed.}

{\ One of the main results of this work is the finding that, for a system with a large momentum relaxation rate where a quasi-hydrodynamic description still applies, the thermo-electric conductivities assume a Drude form when we impose positivity of entropy production and Onsager reciprocity. Specifically, we find that the contribution from the incoherent conductivity to the electric conductivity disappears.}

{\ Typically, the DC conductivity of a material is determined by placing the material on a suitable heat sink and passing a current through it. Once the system settles into a time independent state then one measures the current compared to the applied electric field and obtains the resistance/conductivity. Our work casts serious doubts on whether such measurements can be used to determine incoherent conductivities for situations where we expect hydrodynamics to be a relevant description. Indeed, in work by some of the authors \cite{Brattan:2024dfv}, it now seems likely that the part of the DC conductivity attributed to the incoherent term $\sigma_{0}$ would instead have to originate in some other (combination of) transport coefficients.}

{\ An important step in verifying our theory could be to simulate systems of many charged particles that reach a steady state under the influence of a constant electric field and dissipation, as analyzed from a purely hydrodynamic perspective in this article.  Comparing the theory with numerical experiments could also allow for a detailed investigation into the validity of the hydrodynamic regime around steady states \cite{Brattan:2024dfv}. For example, we can explore the independence of our conclusions from the nature of the microscopic theory by considering interacting exotic particles (e.g.~Lifshitz particles \cite{Brattan:2017yzx,Brattan:2018sgc,Brattan:2018cgk} and anyons \cite{Brattan:2013wya,Brattan:2014moa,Amoretti:2023hpb}). However, the simulations may not be trivial due to the significant number of particles required to achieve a hydrodynamic regime.
Finally, it would be interesting to apply the same techniques to superconductive states along the line of what has been done microscopically in \cite{Solinas:2020woq}.}

{\ We further remark that the focus of this work has been the hydrodynamics of charged fluids dragged by an external electric field, however one could reach similar steady states in the presence of temperature gradients or gravitational fields. Therefore, we could find corrections to the hydrostatic constraint $\partial_i T=0$ due to relaxations similarly to how the constraint \eqref{Eq:ModifiedHydrostaticity} becomes \eqref{Eq:FinalHydrostaticityConstraint} when momentum is relaxed. We remind the reader that our result for the electrical conductivity \eqref{eqn:electric_conductivity} is already very similar to certain expressions obtained using holographic models when the momentum relaxation rate becomes large \cite{Zhou:2015qui,Davison:2015bea,Fu:2022jqn}.}

\acknowledgments

{\noindent A.A. \& D.B. have received support from the project PRIN 2022A8CJP3 by the Italian Ministry of University and Research (MUR). L.M. acknowledges support from the project PRIN 2022ZTPK4E by the Italian Ministry of University and Research (MUR). This project has also received funding from the European Union’s Horizon 2020 research and innovation programme under the Marie Sk\l{}odowska-Curie grant agreement No. 101030915.}

\bibliographystyle{JHEP}
\bibliography{bibliography}

\appendix

\section{Scaling of transport coefficients with spatial velocity}
\label{Appendix:lowvlimit}

{\noindent For completeness we list here the various scalings of our transport coefficients with spatial velocity that are necessary to ensure a smooth $\vec{v} \rightarrow \vec{0}$ limit:
\begin{equation}
	\begin{split}
		\{ \bar{\alpha}_{1,1}, \,\bar{\alpha}_{3,1}, \, \bar{\alpha}_{6,1}, \, \alpha_1 \} & = \mathcal{O}(1), \\
		\{ (\bar{\alpha}_{1,2}-\bar{\alpha}_{1,1}),  (\bar{\alpha}_{3,2}-\bar{\alpha}_{3,1}), \, (\bar{\alpha}_{6,2}-\bar{\alpha}_{6,1}), (\alpha_2 - \alpha_1) \} & = \mathcal{O}(u^2), \\
		\{ \beta_{2,2}, \, \beta_{2,4}, \,\beta_{3,2}, \, \beta_{3,4}, \,\bar{\beta}_{2,2}, \,\bar{\beta}_{2,4}, \bar{\beta}_{5,2}, \,\bar{\beta}_{5,4} \} & = \mathcal{O}(u^2), \\
		\{ (-2 \beta_{2,4} - \beta_{2,2} + \beta_{2,5}), \, (-2 \beta_{3,4} - \beta_{3,2} + \beta_{3,5}) \} & = \mathcal{O}(u^4), \\
		\{ (-2 \bar{\beta}_{2,4} - \bar{\beta}_{2,2} + \bar{\beta}_{2,5}), \, (-2 \bar{\beta}_{5,4} - \bar{\beta}_{5,2} + \bar{\beta}_{5,5}) \} & = \mathcal{O}(u^4), \\
		\mathfrak{t} & = \mathcal{O}(1), \\
		(\gamma_5 - \mathfrak{t}) & = \mathcal{O}(u^2), \\
		\left( 2 \mathfrak{t} + \gamma_2 - \frac{2}{d-1} \mathfrak{t} - 4 \gamma_5 - 2 \gamma_6 + \gamma_3 \right) & = \mathcal{O}(u^4), \\
		\left( -\gamma_3 + \frac{2}{d-1} \mathfrak{t} + \gamma_6 \right) & = \mathcal{O}(u^2), \\
		\left( \gamma_3 - \frac{2}{d-1} \mathfrak{t} \right) & = \mathcal{O}(1).
	\end{split}
\end{equation}
}

\end{document}